\documentclass[aps,prl,reprint,superscriptaddress]{revtex4-2}

\usepackage{bm}
\usepackage{amsmath}
\usepackage[dvipdfmx]{graphicx}
\usepackage{float}
\usepackage{color}

\usepackage {ulem}
\usepackage{cancel}

\graphicspath{{../FIGs/}}

\newcommand{\alphaBETS}{{$\alpha$-(BETS)$_2$I$_3$\ }}
\newcommand{\alphaET}{{$\alpha$-(ET)$_2$I$_3$\ }}

\begin{document}

%\preprint{}

\title{Symmetry-preserving gapped Dirac state revealed by site-selective $^{13}$C NMR in $\alpha$-(BETS)$_2$I$_3$}


\author{Takahiko Sekine}
\affiliation{Department of Applied Physics, University of Tokyo, Bunkyo-ku, Tokyo 113-8656, Japan}
\affiliation{Max-Planck-Institute for Solid State Research, Heisenbergstraße 1, Stuttgart 70569, Germany}
\author{Kazuya Miyagawa}
%\thanks{}
\affiliation{Department of Applied Physics, University of Tokyo, Bunkyo-ku, Tokyo 113-8656, Japan}

\author{Koichi Hiraki}
\affiliation{Integrated Center for Science and Humanities, Fukushima Medical University, Fukushima 960-1295, Japan}
\author{Reizo Kato}
\affiliation{RIKEN, 2-1, Hirosawa, Wako-shi, Saitama 351-0198, Japan}
\author{Kazushi Kanoda}
\affiliation{Department of Applied Physics, University of Tokyo, Bunkyo-ku, Tokyo 113-8656, Japan}
\affiliation{Max-Planck-Institute for Solid State Research, Heisenbergstraße 1, Stuttgart 70569, Germany}
\affiliation{Physics Institute, University of Stuttgart, Pfaffenwaldring 57, 70550 Stuttgart, Germany}

\date{\today}

\begin{abstract}
We report evidence that the low-temperature insulating phase of \alphaBETS preserves both inversion and time-reversal symmetries using site-selective $^{13}$C nuclear magnetic resonance. \alphaBETS undergoes a metal--insulator transition or sharp crossover at approximately 50~K and has been proposed as a candidate spin–orbit-coupling-induced Dirac insulator, yet the nature of its insulating ground state remains unresolved. The NMR spectra exhibit no additional splitting down to 2 K and remain fully described by one quartet and two doublets associated with the three types of molecular sites, A (=A$^{\prime}$), B and C sites, demonstrating the preservation of inversion symmetry. Furthermore, no anomalous line broadening or critical enhancement of the nuclear spin-lattice relaxation rate is observed, indicating the absence of magnetic ordering and time-reversal-symmetry breaking. Analysis of the Knight shifts yields site-specific spin susceptibilities satisfying $\chi_{\mathrm{C}} > \chi_{\mathrm{A}} > \chi_{\mathrm{B}}$, a characteristic feature of tilted Dirac-cone electronic states as observed in $\alpha$-(BEDT-TTF)$_2$I$_3$ under pressure.
However, the faint field-induced ferrimagnetic-like spin excitations observed in $\alpha$-(BEDT-TTF)$_2$I$_3$, which are ascribed to the gapless tilted Dirac cones, are absent in $\alpha$-(BETS)$_2$I$_3$.
These results demonstrate that the insulating ground state of \alphaBETS is a symmetry-preserving gapped Dirac state and provide microscopic constraints on proposed correlated topological phases in this molecular Dirac system.
\end{abstract}

%\keywords{}

\maketitle

\section{I. Introduction}
Massless Dirac fermions in condensed-matter systems have attracted considerable attention as a platform for exploring unconventional electronic states emerging from the interplay of symmetry, topology, and electron correlations. In addition to graphene \cite{Novoselov2004,Novoselov2005}, molecular conductors have become an important class of Dirac materials owing to their highly tunable electronic structures and strong Coulomb interactions. Among them, the organic conductor $\alpha$-(BEDT-TTF)$_2$I$_3$ (BEDT-TTF = bis(ethylenedithio)-tetrathiafulvalene, abbreviated as ET hereafter) has been extensively studied as a prototype correlated Dirac-fermion system \cite{Hirata2016,Hirata2017,Hirata2021,Tajima2006,Katayama2006}. Under pressure, \alphaET exhibits two-dimensional tilted Dirac cones at the Fermi level and a variety of interaction-driven phenomena; velocity renormalization, ferrimagnetic-like spin excitations \cite{Hirata2016}, and anomalous spin fluctuations \cite{Hirata2017, Ohki2019}. More recently, interlayer coupling has been argued to induce Weyl semimetal states in this system \cite{Osada2024,Osada2025}.

On the other hand at ambient pressure, \alphaET undergoes a metal–insulator transition near 135~K \cite{Tajima2006}. Extensive experimental and theoretical studies have established that the insulating phase is a charge-ordered state accompanied by inversion-symmetry breaking \cite{Kakiuchi2007}. Site-selective nuclear magnetic resonance (NMR) measurements played a central role in identifying this state by directly revealing the inequivalence of molecular sites in the unit cell below the transition temperature \cite{Kawai2009,Ishikawa2016}. The contrast between the pressure-induced Dirac-fermion phase and the charge-ordered insulating phase highlights the strong sensitivity of molecular Dirac systems to competing electronic instabilities.

\begin{figure*}
\centering
\includegraphics[width = 15cm]{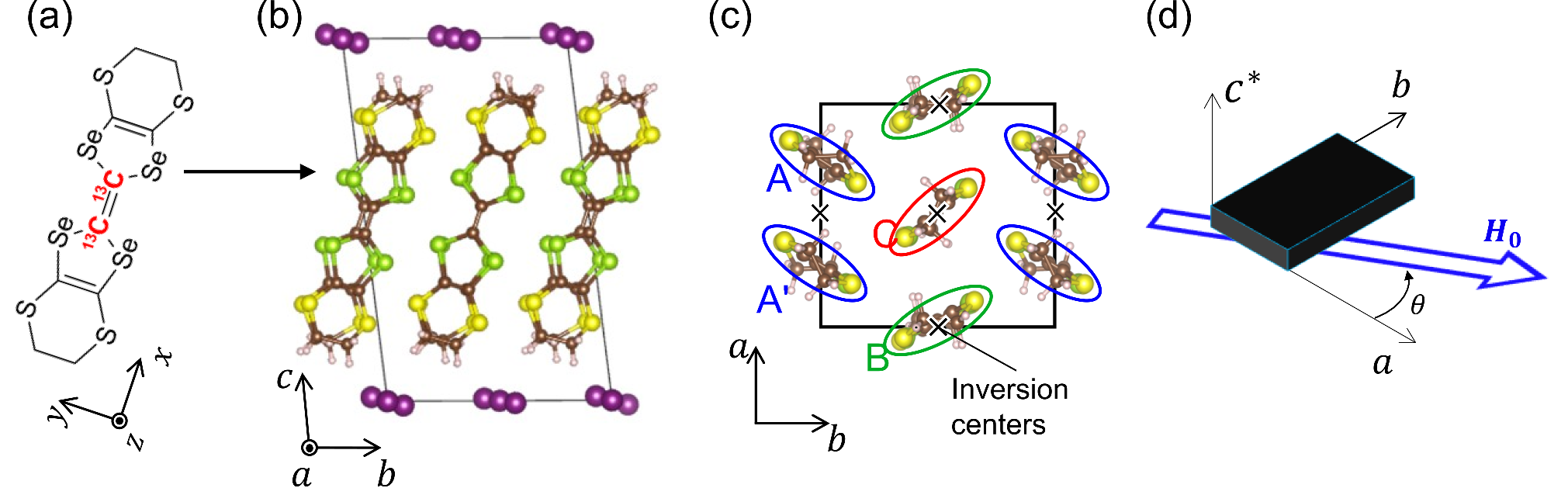}
\caption{\label{Fig1}(a) BETS molecule whose central double-bonded carbons are enriched by $^{13}$C isotopes. (b) Crystal structure of $\alpha$-(BETS)$_2$I$_3$, where BETS molecule layers and I$_3$ layers are alternatively stack. (c) BETS molecule coordinate in the unit cell, which four molecular sites, A, A$^\prime$, B, and C. Inversion centers are located on the B and C sites and between A and A$^\prime$ sites. Thus, A and A$^\prime$ sites are equivalent. (d) Field orientation against a plate-shape single crystal of $\alpha$-(BETS)$_2$I$_3$, which is applied within $ab$ layer.}
\end{figure*}

The selenium analog of $\alpha$-(ET)$_2$I$_3$, $\alpha$-(BETS)$_2$I$_3$, provides a unique opportunity to investigate a different route toward an insulating state in a molecular Dirac system \cite{Kato1991,Inokuchi1993,Inokuchi1995,Kobayashi2004a} Because the BETS molecule contains four selenium atoms instead of the inner sulfur atoms, spin–orbit coupling is substantially enhanced compared with that in $\alpha$-(ET)$_2$I$_3$ \cite{Winter2017}. Band-structure calculations have predicted that \alphaBETS possesses Dirac-like electronic bands near the Fermi energy and that spin–orbit coupling can open a gap at the Dirac points \cite{Tsumuraya2021,Kondo2009}. Furthermore, recent theoretical studies suggested that electron correlations may cooperate with spin–orbit coupling to stabilize topologically nontrivial insulating states, including correlated quantum spin Hall phases \cite{Raghu2008, Ohki2022}.

Experimentally, \alphaBETS also exhibits a metal–insulator transition or sharp crossover near 50~K at ambient pressure \cite{Inokuchi1995}. In contrast to the established charge-ordered state in $\alpha$-(ET)$_2$I$_3$, the nature of the insulating phase remains controversial. Transport measurements indicate the opening of an energy gap, whereas magnetic torque measurements have been interpreted as evidence for unconventional low-temperature magnetic behavior \cite{Konoike2022}. The central unresolved question is therefore whether the insulating phase arises from spontaneous symmetry breaking or  remains a symmetry-preserving gapped Dirac state. Determining the fate of inversion symmetry and time-reversal symmetry is essential for distinguishing between these possibilities.

NMR is particularly well suited to addressing this question because it directly probes local electronic environments and internal magnetic fields on the molecular scale.
Spectral splitting can straightforward reveal the emergence of inequivalent molecular sites with inversion-symmetry breaking \cite{Miyagawa2000,Ishikawa2016}, while internal magnetic fields generated by magnetic order with time-reversal-symmetry-breaking states can be detected through line splitting, anomalous line broadening, and an enhancement of the nuclear spin-lattice relaxation rate. In addition, site-selective NMR provides access to local spin susceptibilities, allowing a microscopic investigation of the electronic structure of Dirac-cone bands \cite{Hirata2016}.

In this work, we perform site-selective $^{13}$C-NMR measurements on single-crystalline \alphaBETS over a wide temperature range spanning the metal–insulator transition. By analyzing the angular dependence of the spectra and the Knight shifts, we estimate the local spin susceptibilities at the three inequivalent molecular sites. We further investigate low-energy spin dynamics by measuring the nuclear spin-lattice relaxation rate. Our results demonstrate that the low-temperature insulating phase preserves both inversion and time-reversal symmetries down to at least 2 K. The site-specific spin susceptibilities exhibit the characteristic relation $\chi_{\mathrm{C}} > \chi_{\mathrm{A}} > \chi_{\mathrm{B}}$, as observed in tilted Dirac cones of \alphaET \cite{Hirata2016}. At the same time, the absence of magnetic and charge ordering indicates that the low-temperature phase is a symmetry-preserving gapped Dirac state. These findings provide microscopic constraints on theoretical descriptions of \alphaBETS and clarify the nature of its insulating ground state.

\section{II. Experimental and Analysis}
\subsection{A. Sample preparatoin and NMR measurements}
Single crystals of \alphaBETS enriched with $^{13}$C isotopes at the central double-bonded carbon sites were prepared by the electrochemical oxidation method. Figure~\ref{Fig1}(a) shows the molecular structure of BETS and the positions of the $^{13}$C-enriched sites. The crystal structure is triclinic with $P\bar{1}$ space group at ambient temperature~\cite{Kato1991,Kitou2021}, consisting of alternating BETS and I$_{3}$ layers (Figure~\ref{Fig1}(b)).
The unit cell contains four BETS molecules denoted A, A$^\prime$, B, and C (Fig.~\ref{Fig1}(c)), where A and A$^\prime$ molecular sites are equivalent due to inversion.
$^{13}$C-NMR measurements were performed using a phase-coherent pulsed NMR spectrometer under a magnetic field of 6.0~T. The magnetic field was rotated within the conducting ab plane using a split-pair superconducting magnet equipped with a mechanical rotation stage. The field direction is specified by the angle $\theta$ measured from the crystallographic a axis (Fig.~\ref{Fig1}(d)). NMR spectra were acquired using spin-echo and solid-echo pulse sequences \cite{Powles1962}.
Both methods yielded identical spectral profiles within experimental accuracy. The NMR frequency was calibrated using tetramethylsilane (TMS) as a reference. The nuclear spin-lattice relaxation rate $1/T_1$ was measured using the saturation-recovery method. Recovery curves were recorded at a fixed field orientation $(\theta = 31^\circ )$ and analyzed as described below.

\begin{figure}[b]
\centering
\includegraphics[width = \columnwidth]{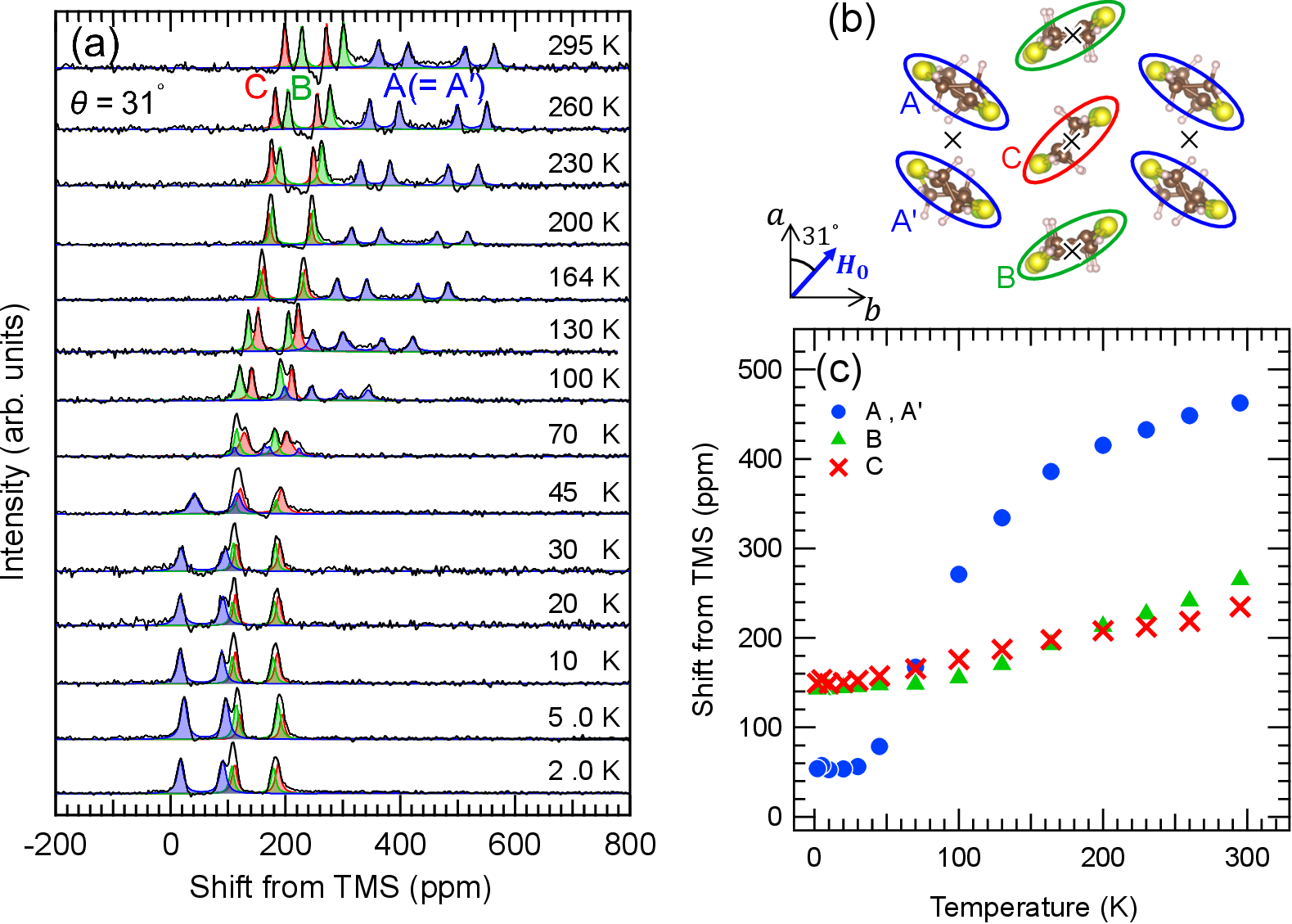}
\caption{\label{Fig2}$^{13}$C NMR spectra under the field at $\theta = 31^\circ$. Lorentzian line-shape quartet function (see the text) for the A (=A$^\prime$) site is highlighted by blue. The doublet functions for the B and C sites are highlighted by green and red, respectively. (b) Field orientation in the unit cell at $\theta = 31^\circ$, which is close to perpendicular to A and A$^\prime$ sites. (c) Central-frequency spectral shifts for each site.}
\end{figure}

\begin{figure}[b]
\centering
\includegraphics[width = \columnwidth]{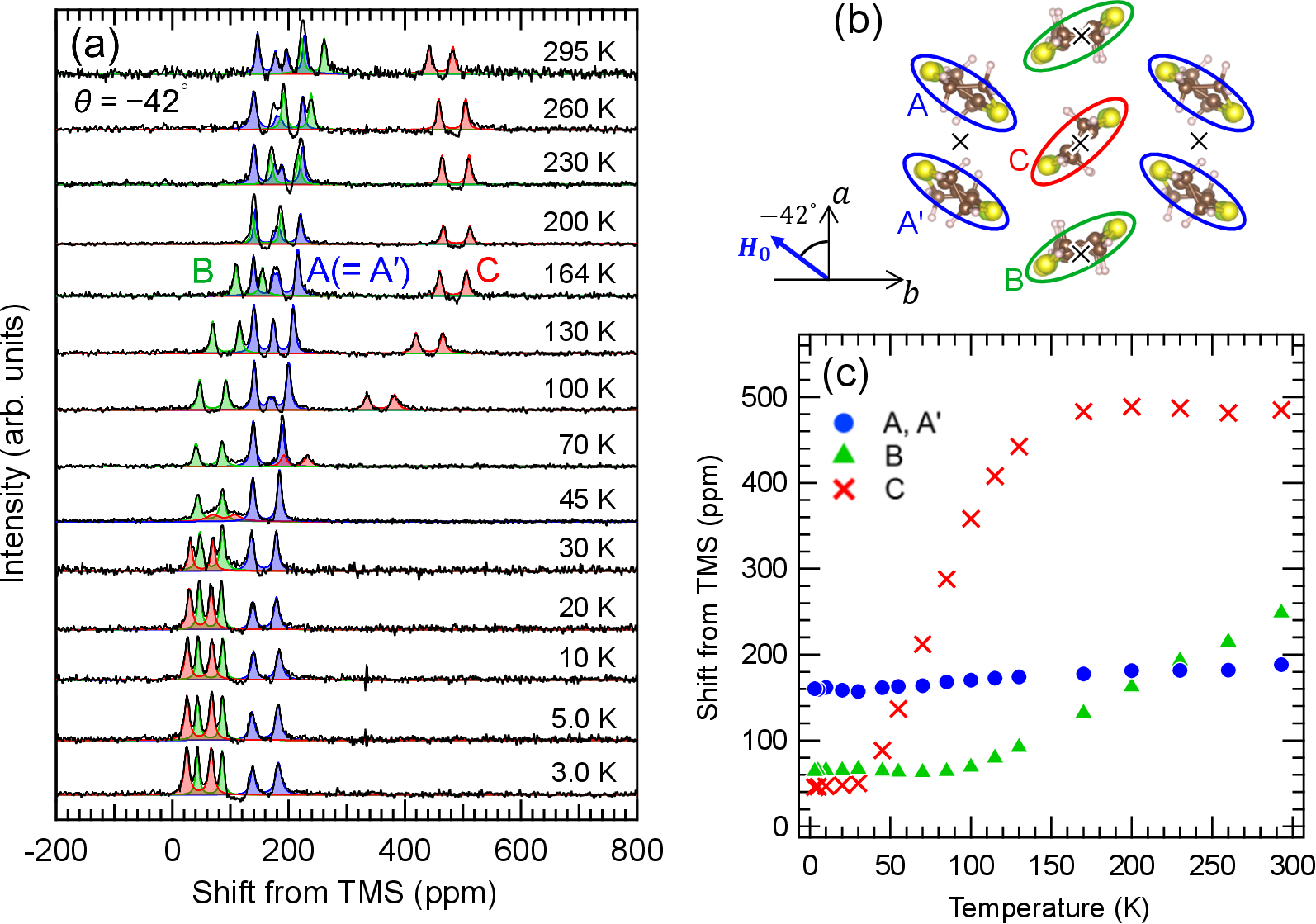}
\caption{\label{Fig3}(a) $^{13}$C NMR spectra under the field at $\theta = -42^\circ$. Legends of each resonant peak are the same as the Fig.~\ref{Fig2}(a). (b) Field orientation at $\theta = -42^\circ$, which is close to perpendicular to B and C sites. (c) Central-frequency spectral shifts for each site.}
\end{figure}

\subsection{B. Analysis of NMR spectra}
The two central $^{13}$C nuclei in a BETS molecule are coupled through the nuclear dipole–dipole interaction \cite{Kawamoto1995}. Owing to the crystal symmetry, the paired $^{13}$C nuclei are equivalent within each B and C molecular site whereas they are inequivalent within each A and A$^\prime$ molecular site.
For equivalent nuclei, the dipolar-coupled two-spin system produces a Pake doublet.
In contrast, inequivalent nuclei give rise to a quartet  consisting of four allowed transitions between the coupled spin states \cite{Hirata2011}. Consequently, the NMR spectrum of \alphaBETS consists of one quartet originating from the A (=A$^\prime$) molecular site and two doublets arising from the B and C molecular sites.
The spectra were fitted using Lorentzian line-shape functions derived from the dipolar-coupled two-spin model. The B- and C-site spectra were represented by doublet functions, whereas the A-site spectrum was represented by a quartet function. The fitting parameters were the center frequency, linewidth, spectral intensity, and inequivalence parameter for the paired $^{13}$C on the A site.
The hyperfine coupling tensor of the $^{13}$C spin is dominated by the anisotropic $p_z$ orbital contribution and can be approximated by a uniaxial tensor,
\begin{equation}
\hat{A}=
\begin{pmatrix}
A_{xx} &&\\ &A_{yy} &\\ &&A_{zz} 
\end{pmatrix}
=A_{\mathrm{iso}} \hat{I} + A_{\mathrm{aniso}}
\begin{pmatrix}
-1 &&\\ &-1 &\\ &&2 
\end{pmatrix}
\end{equation}
where $A_{\mathrm{iso}}$ and $A_{\mathrm{aniso}}$ denote the isotropic and anisotropic hyperfine coupling constants, respectively.
We note that this tensor is usually diagonalized by the molecular principal axes, $(x, y, z)$, whose orientations are independent of the lattice axes, $(a, b, c)$ (Fig.~\ref{Fig1}(a) and \ref{Fig1}(b)): The $x$ and $y$ axes correspond to the long and short axes of the molecular plane, respectively, and the $z$ axis is perpendicular to the plane (Fig.~\ref{Fig1}(a)).
Because of the anisotropy of the tensor, the Knight shifts strongly depend on the field orientation. The different molecular orientations of the A, B, and C sites in the unit cell enable site-selective detection of local spin susceptibilities via the angular dependence of the NMR spectra at each site.

\subsection{C. Analysis of nuclear spin-lattice relaxation rate}
The relaxation curves were analyzed using both double-exponential and stretched-exponential functions. In principle, $1/T_1$ depends on the transverse component of the hyperfine coupling tensor with respect to the applied magnetic field \cite{Moriya1963}. At $\theta = 31^\circ$, the anisotropy of the hyperfine coupling tensor results in a relatively small transverse hyperfine field at the A and A$^\prime$ sites and a larger transverse hyperfine field at the B and C sites. Consequently, the relaxation curves are described by the sum of two characteristic relaxation components, $1/T_1^{\mathrm{long}}$ and $1/T_1^{\mathrm{short}}$, predominantly associated with the A/A$^\prime$ and B/C sites, respectively; $1-M(t)/M(\infty)=A^{\mathrm{long}} \exp\left(-t⁄T_1^{\mathrm{long}}\right)+A^{\mathrm{short}} \exp\left(-t⁄T_1^{\mathrm{short}}\right)$. To characterize the overall relaxation behavior, the relaxation curves were also analyzed using a stretched-exponential function, $1-M(t)/M(\infty)=\exp\left[-\left(t/T_1\right)^{\beta}\right]$, where $\beta$ represents the distribution of relaxation rates. The parameter $\beta$ approaches unity when the relaxation is spatially homogeneous and decreases as the distribution of local relaxation rates broadens. The effective relaxation rate estimated from the stretched-exponential analysis is used below to discuss the low-energy spin dynamics.

\begin{figure}[t]
\centering
\includegraphics[width = .7\columnwidth]{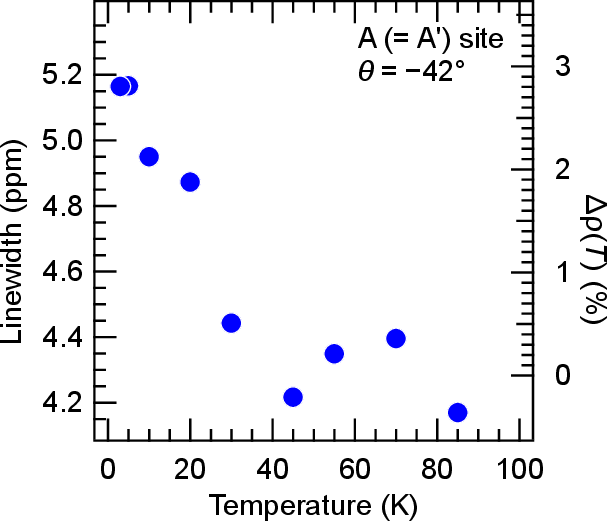}
\caption{\label{Fig3_2}Linewidth and possible charge difference from 45--55 K estimated by using the chemical shift tensor of \alphaET\cite{Hirata2011}. }
\end{figure}

\begin{figure*}[t]
\centering
\includegraphics[width = 1.8\columnwidth]{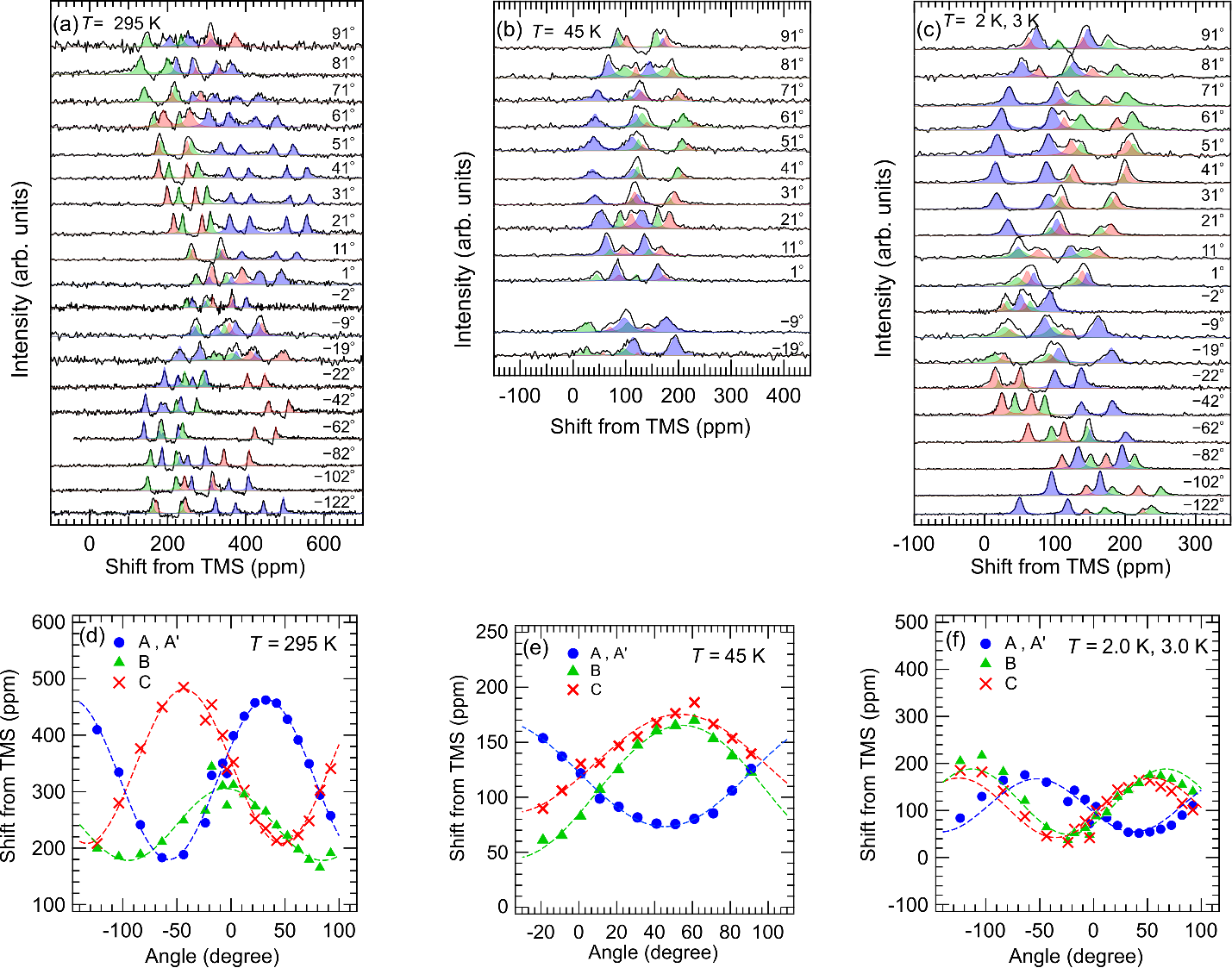}
\caption{\label{Fig4}NMR spectra under various field orientations within $ab$ plane at (a)~295~K, (b)~45~K, and (c)~2--3~K. Blue areas show the Lorentzian line-shape quartet functions for A(=A$^\prime$) site and green and red areas show those doublet functions for B and C sites, respectively, in the same way as Fig.~\ref{Fig2}(a) and \ref{Fig3}(a). Field-angle dependences of central-frequency shifts at (d)~295~K, (e)~45~K and (f)~2--3~K.}
\end{figure*}

\section{III. Results}
\subsection{A. Temperature evolution of the $^{13}$C-NMR spectra}
Figure~\ref{Fig2}(a) shows the temperature evolution of the $^{13}$C-NMR spectra measured with the magnetic field applied at $\theta = 31^\circ$ (Fig.~\ref{Fig2}(b)). The spectra are well fitted by the functions described in Sec.~II~B, as shown in the figure, and assigned to the A (= A$^\prime$), B, and C molecular sites. At ambient temperature, eight resonance peaks are clearly resolved into one quartet and two doublets. Upon cooling, the spectral shift of the A site continuously moves toward lower frequency. Its quartet structure gradually evolves into a doublet-like profile. In contrast, the B- and C-site spectral shifts exhibit substantially weaker  temperature dependence under this field orientation.
The site-selective center-frequency shifts obtained from the fitting analysis are summarized in Fig.~\ref{Fig2}(c). All shifts continuously decrease upon cooling and exhibit neither detectable discontinuity nor anomaly around the metal–insulator transition temperature, $T_{\mathrm{MI}} \sim 50~$K.
Furthermore, no additional resonance lines appear at any temperature. The linewidths remain nearly constant across $T_{\mathrm{MI}}$ without an abrupt broadening indicative of the emergence of static internal magnetic fields.

To further examine the site dependence of the spectral shifts, we also measured the spectra at another field orientation with $\theta = -42^\circ$, where the B- and C-site shifts are more pronounced at ambient temperature (Fig.~\ref{Fig3}(a) and \ref{Fig3}(b)).
The spectra remain well resolved over the entire temperature range. For this field orientation, the B- and C-site shifts exhibit stronger temperature dependence than the A-site shift. The estimated center-frequency shifts are summarized in Fig.~\ref{Fig3}(c).

As seen above, there are neither of spectral splitting nor broadening signifying magnetic order or symmetry breaking between the A and A$^\prime$ molecular sites.
Below, we look into the spectra in more detail.
With $\theta = -42^\circ$, the A $(=\rm A^\prime$) site shift are almost independent of temperature (Fig.~\ref{Fig3}(c)).
This indicates that the shift and line profiles are dominated by the chemical shift for this field orientation.
Figure~4 shows the temperature dependence of the linewidth of the A/A$^\prime$-site spectra below 100~K. The linewidth is quite sharp (of the order of ppm) and only increases by $\sim$1 ppm upon cooling from 50 K.
Although the broadening to this extent can be understood as resulting from imperfections in the crystal, we estimate the upper limit of the charge disproportionation, if any.
We estimate the possibile charge difference between the A and A$^\prime$ sites, $\Delta \rho^{\rm A/A'}$, using the relationship between the local molecular valence $\rho$ and the chemical shift known for $\alpha$-(ET)$_2$I$_3$.
The chemical-shift tensor of \alphaBETS is expected to be very similar to that of \alphaET because the molecular wave functions, which primarily determine the chemical-shift tensor, are expected to be similar for BETS and ET.
In $\alpha$-(ET)$_2$I$_3$, the chemical-shift-tensor components are $\sigma_{xx}=130.8\rho+46.2$, $\sigma_{yy}=30.4\rho+159.0$, and $\sigma_{zz}=-12.1\rho+61.9$ (ppm)\cite{Hirata2011}.
For the A site of $\alpha$-(BETS)$_2$I$_3$, the chemical shift at $\theta = -42^{\circ}$ is  $\sigma^{\rm A/A'}=31.4 \rho + 158.1$.
Thus, $\Delta \rho^{\rm A/A'}$ can be estimated as $\Delta \omega^{\rm A/A'}/31.4$, where $\Delta\omega^{\rm A/A'}$ is an additional linewidth below $T_{\rm MI}$ in Fig.~\ref{Fig3_2}.
Figure~\ref{Fig3_2} also displays the etimated $\Delta\rho^{\rm A/A'}$ values, which reach less than 3\%. This is the upper limit of charge disproportionation if the slight line broadening is assumed to solely originate from the charge imbalance.

Again, all spectral parameters continuously evolve across $T_{\rm MI}$ without any detectable anomaly.
These observations demonstrate that the local electronic environments continuously evolve through the metal–insulator transition and remain well described by the same spectral model over the entire temperature range.

\subsection{B. Angular dependence of the spectra}
To investigate the symmetry more precisely, we measured the field-angle dependence of the spectra while rotating the magnetic field within the conducting ab plane. The different orientations of A (=A$^\prime$), B, and C molecules in the unit cell give distinct field-angle dependencies of their spectra.
Figures~\ref{Fig4}(a)--\ref{Fig4}(c) show the spectra measured at 295~K, 45~K, and 2--3~K, respectively.
{\color{black}In Fig~\ref{Fig4}(c), the spectra were measured at 2~K for $\theta \geq -9^\circ$ and at 3~K for  $\theta \leq -19^\circ$. As shown in Figs.~\ref{Fig2}(c) and \ref{Fig3}(c), no spectral shifts occur between 2 and 3~K.
Therefore, the spectral profiles are expected to be indentical at these temperatures. For simplicity, we describe these spectra as being at 2~K in the following discussion.}
At all temperatures, the spectra are well-fitted by one quartet and two doublets. Even below $T_{\mathrm{MI}}$, no additional resonance peaks appear and no anomalous spectral broadening or splitting is observed.
The linewidths display only weak angular dependence.
In particular, the spectral features associated with the A and A$^\prime$ sites remain equivalent within experimental resolution at both 45~K and 2~K.  Indeed, the site-selective center-frequency shifts plotted in Figs.~\ref{Fig4}(d)--\ref{Fig4}(f) exhibit smooth angular variation and remain well resolved over the entire angular range. These results indicate that the equivalence between A and A$^\prime$ sites is preserved down to 2~K.
The angular dependence of the resonance lines strongly temperature-dependent because of the interplay between the strongly temperature-dependent Kinight shift tensor and temperature-independent chemical shift tensor, as seen below.

\begin{figure}[b]
\centering
\includegraphics[width = .8\columnwidth]{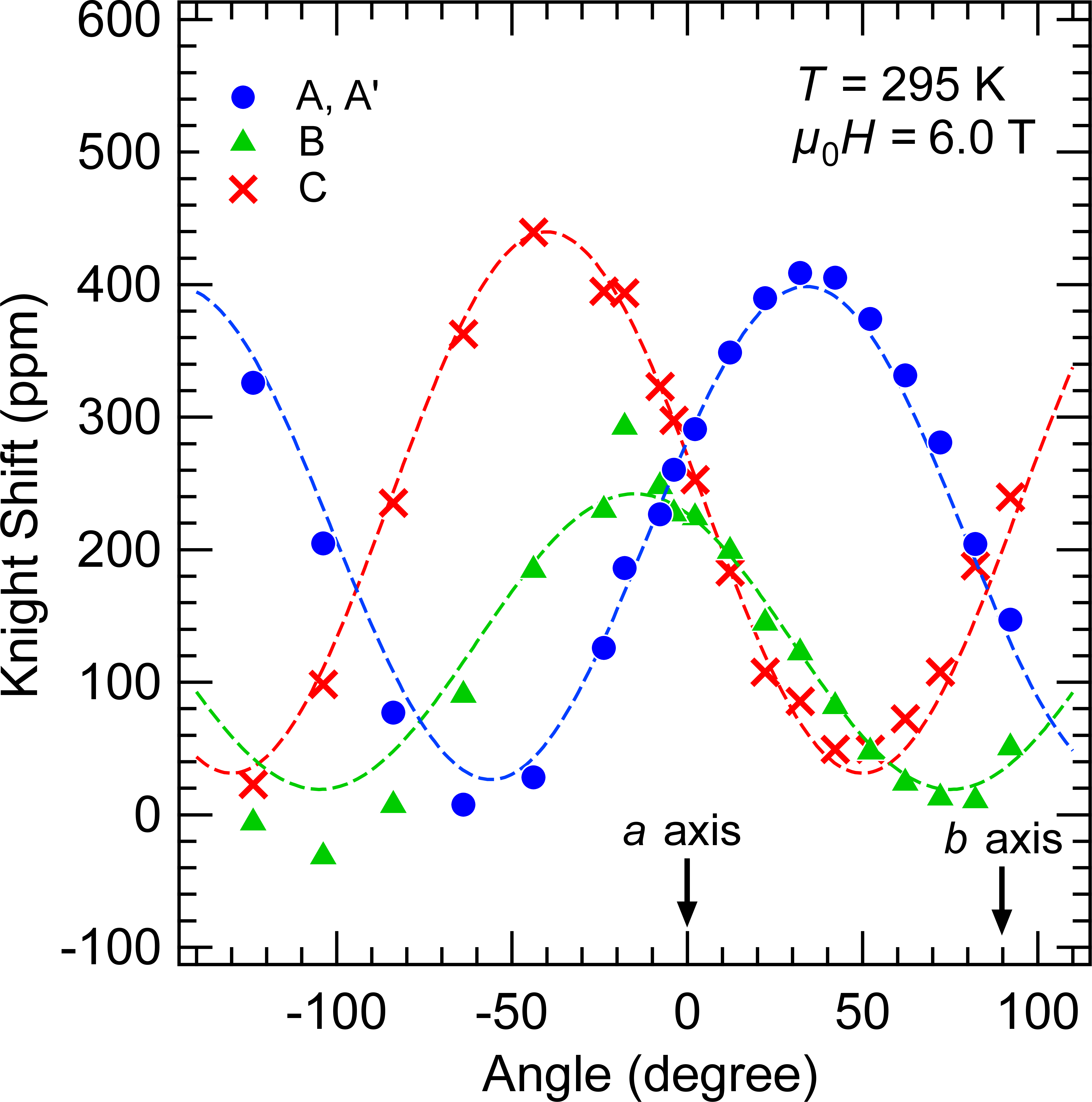}
\caption{\label{Fig5}(a) Field-dependence of the Knight shifts from the A/A$^\prime$ (blue), B (green), and C (red) sites. Dotted lines are the fitting curve using uniaxial anisotropic Knight-shift tensor as explained in the text.}
\end{figure}

\subsection{C. Knight-shift tensor analysis}
To determine the site-selective spin susceptibility, the hyperfine-coupling tensor was first evaluated from the angular dependence of the Knight shifts. As described in Sec.~II~B, the hyperfine interaction at the central $^{13}$C nuclei is approximated by a uniaxial tensor $\hat{K}_\mu$. The Knight shift is therefore expressed as
\begin{equation}
K_\mu(\theta) = \frac{\bm{H}(\theta)\hat{K}_\mu \bm{H}(\theta)}{H_0^2} \ (\mu = \mathrm{A, B, C}),
\end{equation}
where $\bm{H}(\theta)$ is the magnetic field applied at an angle $\theta$ in the $ab$ plane as shown in Fig.~\ref{Fig4}(d), i.e., $\bm{H}(\theta))⁄H_0 =\bm{a}\cos\theta+\bm{b}^\prime\sin\theta$, where $\bm{b}^\prime$ is an orthogonalized crystal axis. Since the principal axes of the Knight shift tensor, $(\bm{x}^\mu, \bm{y}^\mu, \bm{z}^\mu)$, are irrelevant to the crystallographic axes, $(\bm{a}, \bm{b}^\prime, \bm{c}^*)$, we introduce the coordinate-transformation matrix $\hat{T}_\mu$, which satisfies $\hat{T}_\mu (\bm{a}, \bm{b}^\prime, \bm{c}^* )=(\bm{x}^\mu, \bm{y}^\mu, \bm{z}^\mu)$. Thus, the Knight shift $\hat{K}_\mu$ is rewritten as
\begin{equation}
K_\mu(\theta) = \frac{\bm{H}(\theta-\theta_0^\mu)\hat{T}_\mu^{-1}\hat{K}_\mu\hat{T}_\mu \bm{H}(\theta-\theta_0^\mu)}{H_0^2}, \ (\mu = \mathrm{A, B, C}).
\end{equation}
$\theta_0^\mu$ is an angular offset introduced to compensate for the misalignment between the principal axes in hyperfiine-coupling tensor and the molecular-plane axes as shown in Fig.~\ref{Fig1}(a). If these two sets of axes are exactly aligned, $\theta_0^\mu = 0$. Otherwise a finite value of $\theta_0^\mu$ is required to fit the experimental values using Eq~(3). 

To extract the Knight shifts from the experimental data in Fig.~\ref{Fig4}(d)--\ref{Fig4}(f), the spectra at 2~K were used as a reference for the chemical shifts. Because the spin susceptibility is strongly suppressed and the spectral shifts stay unchanged at low temperatures, the residual shifts at 2~K are solely attributable to the chemical shifts.
The Knight shifts at ambient temperature were therefore estimated from the difference between the shifts at 295~K and 2~K.
Figure~\ref{Fig5} shows the angular dependence of the Knight shifts for the A, B, and C sites. The data exhibit clear sinusoidal behavior and are well fitted by the above expression.

The fitting parameters are summarized in Table I. The ratios $K_{\mathrm{iso}}/K_{\mathrm{aniso}}$ are nearly identical for all three sites, which suggests that the hyperfine coupling constants are approximately independent of the molecular sites. This is reasonable because the hyperfine coupling constant is primarily determined by the molecular orbital and not largely site-dependent. The obtained values of $\theta_0^\mu$ are close to zero for the A and C sites, indicating that those principal axes of the hyperfine-coupling tensors are almost aligned with the corresponding molecular structure.
However, small but finite deviations are observed for the B site, suggesting the presence of additional contributions to the local hyperfine field that cause the misalignment between the principal axes of the hyperfine coupling tensor and the symmetry axes of the molecular structure.
One possible contribution is inter-site spin--spin interactions. Although such a van-der Waals crystal has usually small inter-site interactions, exchange fields from A/A$^\prime$ and C sites to B site might be relatively larger since the B-site Knight shift  is smaller than the others.

\begin{figure}
\centering
\includegraphics[width = 1.03\columnwidth]{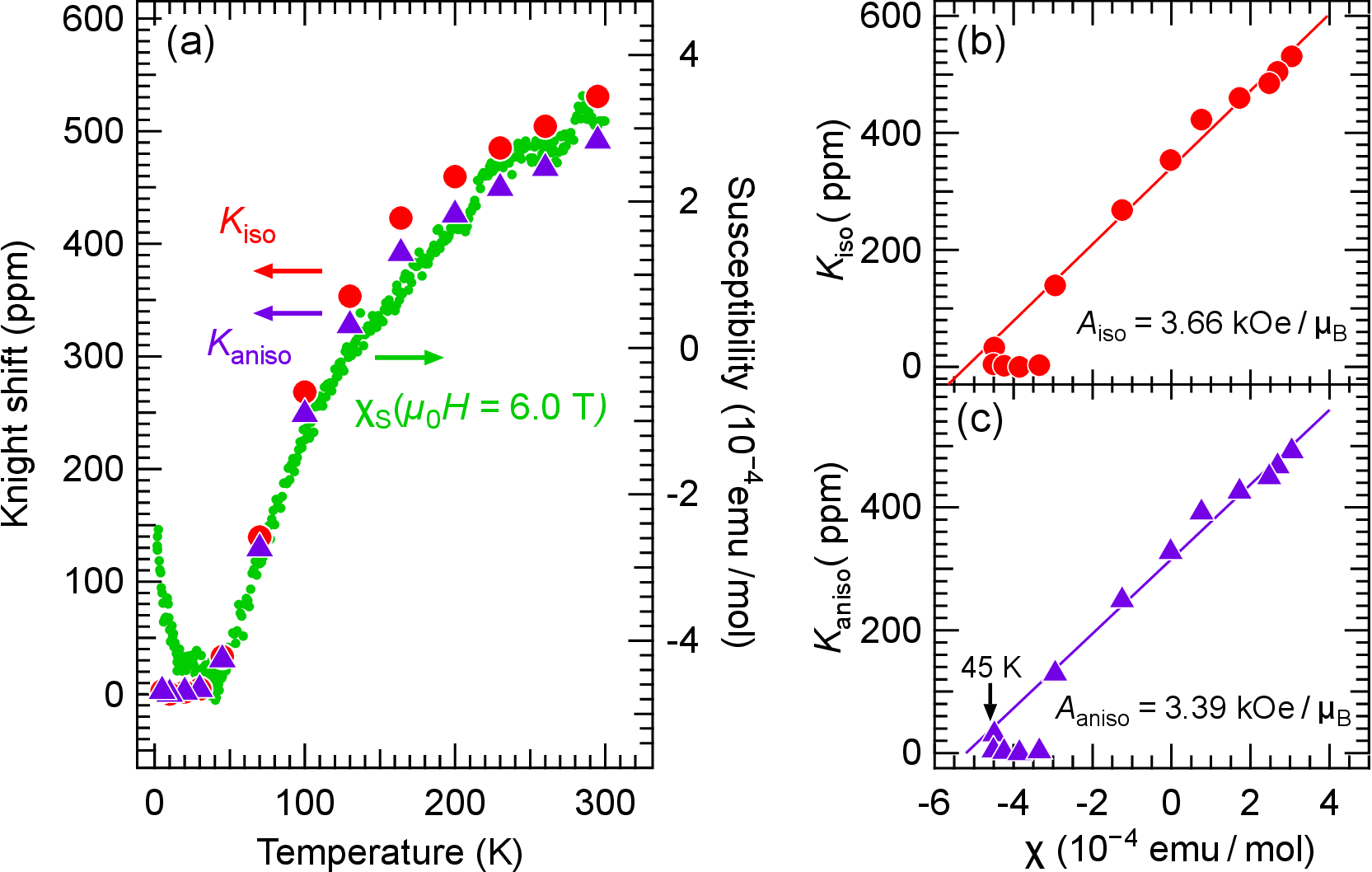}
\caption{\label{Fig6}(a) Temperature dependence of $K_\mathrm{iso}$, $K_\mathrm{aniso}$ and macroscopic susceptibility $\chi$. $K_\mathrm{iso}$ and $K_\mathrm{aniso}$ are estimated by using the Knight-shift tensors determined by the fitting in Fig.~\ref{Fig5}(b) and \ref{Fig5}(c) show $K$--$\chi$ plots using $K_\mathrm{iso}$ and $K_\mathrm{aniso}$, respectively. Linear fittings above 45~K give the hyperfine coupling constants of $K_\mathrm{iso} = 3.66 $~kOe/$\mu_\mathrm{B}$ and $K_\mathrm{aniso}=3.39$~kOe/$\mu_{\mathrm{B}}$ (see the text).}
\end{figure}

\begin{table}[b]
\centering
\begin{ruledtabular}
\begin{tabular}{c|lll}
Fitting parameter & A/A$^\prime$& B &C\\\hline
$K_{\mathrm{iso}}$ & 141 & 86.3 & 153\\
$K_{\mathrm{aniso}}$ & 131 & 79.2 &144\\
$K_{\mathrm{aniso}}/K_{\mathrm{iso}}$ &0.93&0.92&0.93\\
$\theta_0$&$-0.16$&$-12$&1.6 \\
\end{tabular}
\end{ruledtabular}
\caption{Fitting parameters for the field-angle dependence of the Knight shifts}
\end{table}
Using the obtained tensor parameters, we next estimate the temperature dependence of the Knight shifts at fixed field orientations. For the A (=A$^\prime$) site, the spectra at $\theta = 31^\circ$ were used, whereas those at $\theta = -42^\circ$ were used for the B and C sites, where each hyperfine coupling constant is sufficiently large. Substituting the fitting parameters in Table~I into Eq. (2), the observed Knight shifts are expressed as
\begin{eqnarray}
K^{\mathrm{A/A^{\prime}}}(\theta =31^\circ)&=& K^{\mathrm{A/A^{\prime}}}_{\mathrm{iso}}\left(1 + 1.96\cdot \left[K^{\mathrm{A/A^{\prime}}}_{\mathrm{aniso}}/K^{\mathrm{A/A^{\prime}}}_{\mathrm{iso}}\right]\right) \notag\\
K^{\mathrm{B}}(\theta =-42^\circ)&=& K^{\mathrm{B}}_{\mathrm{iso}}\left(1 + 1.30 \cdot\left[K^{\mathrm{B}}_{\mathrm{aniso}}/K^{\mathrm{B}}_{\mathrm{iso}}\right]\right) \\
K^{\mathrm{C}}(\theta =-42^\circ)&=& K^{\mathrm{C}}_{\mathrm{iso}}\left(1 + 1.95\cdot \left[K^{\mathrm{C}}_{\mathrm{aniso}}/K^{\mathrm{C}}_{\mathrm{iso}}\right]\right) \notag
\end{eqnarray}
Assuming that these equations hold at every temprature, the temperature dependence of $K^\mu_{\mathrm{iso}} \ (\mu = \mathrm{A, B, C})$ are given as follows:
\begin{eqnarray}
K_{\mathrm{iso}}^{\mathrm{A/A^\prime}}(T) &=& \left[K^{\mathrm{A/A^\prime}}(T;\theta =31^\circ)\right]/(1 + 1.96\cdot 0.93) \notag\\
K_{\mathrm{iso}}^{\mathrm{B}}(T) &=& \left[K^{\mathrm{B}}(T;\theta =-42^\circ)\right]/(1 + 1.30\cdot 0.92) \\
K_{\mathrm{iso}}^{\mathrm{C}}(T) &=& \left[K^{\mathrm{C}}(T;\theta =-42^\circ)\right]/(1 + 1.95\cdot 0.93) 
\end{eqnarray}
The temperature dependence of $K_{\mathrm{aniso}}^\mu \ (\mu=\mathrm{A,B,C})$ is given through the relation of $(K_{\mathrm{aniso}}^\mu)/(K_{\mathrm{iso}}^\mu)=0.93 (\mu=\mathrm{A/A^\prime})$  and 0.92 $(\mu=\mathrm{B,C})$.
Figure~\ref{Fig6}(a) displays $K_{\mathrm{iso}}^\mu(T)$ and $K_{\mathrm{aniso}}^\mu(T)$. Both components continuously decrease upon cooling and become nearly independent of temperature below approximately 30~K.
To determine the hyperfine coupling constants $A_{\mathrm{iso(aniso)}}^\mu$, the bulk magnetic susceptibility, $\chi$, measured using a Quantum Design MPMS3 SQUID magnetometer, is compared with the total Knight shift summed over the unit cell, $K_{\mathrm{iso}}^{\mathrm{total}}(T)=2K_{\mathrm{iso(aniso)}}^{\mathrm{A}} (T)+K_{\mathrm{iso(aniso)}}^{\mathrm{B}} (T)+K_{\mathrm{iso(aniso)}}^{\mathrm{C}}(T)$.
Figures~\ref{Fig6}(b) and~\ref{Fig6}(c) show the corresponding $K$--$\chi$ plots, where the data of $\chi$ measured at 6~T was used. Linear relationships, $K_{\mathrm{iso(aniso)}}^\mathrm{total}=A_\mathrm{iso(aniso)}\chi(T)+const.$, are observed above approximately 45~K, whereas deviations appear at lower temperatures.
The linear fits yield $A_{\mathrm{iso}} = 3.66~\mathrm{kOe}/\mu_{\mathrm{B}}$ and $A_{\mathrm{aniso}} = 3.39~\mathrm{kOe}/\mu_{\mathrm{B}}$. These hyperfine coupling constants are comparable to those reported for $\alpha$-(ET)$_2$I$_3$, for which $A_{\mathrm{iso}} = 3.54~\mathrm{kOe}/\mu_{\mathrm{B}}$ and $A_{\mathrm{aniso}} = 4.82~\mathrm{kOe}/\mu_{\mathrm{B}}$\cite{Tajima2006}.
While the isotropic component is similar between the two compounds, the anisotropic component is substantially reduced in \alphaBETS. This difference may originate from the distinct molecular-orbital profiles resulting from the substitution of S adjacent to the central $^{13}$C by Se.

\begin{figure*}
\centering
\includegraphics[width = 1.65\columnwidth]{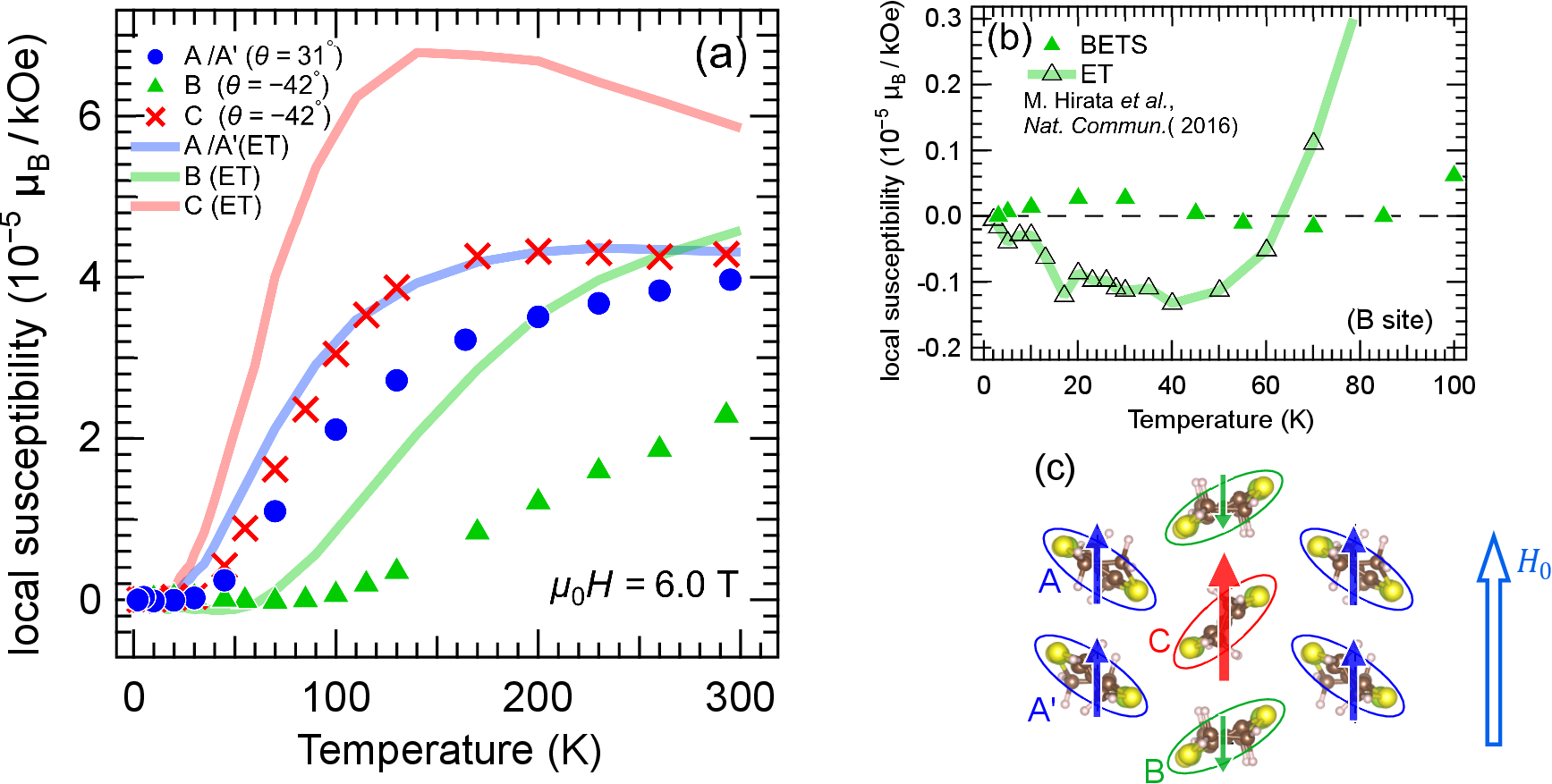}
\caption{\label{Fig7}(a) Local susceptibilities from each site in \alphaBETS estimated by using the hyperfine coupling constants determined in Fig.~\ref{Fig6}(b) and \ref{Fig6}(c) (see the text). Bold lines show the susceptibilities of \alphaET under pressure of 2.3~GPa \cite{Hirata2016}. The low-temperature B-site susceptibilities are focused in (b). \alphaET shows negative values below 60~K, which suggests ferrimagnetic spin excitations as in (c).}
\end{figure*}

%{\color{blue}Furthermore, the present susceptibility becomes negative below approximately 150~K under a field of 6~T, reaching a minimum value of about $-5 \times 10^{-4} emu mol^{-1}$. The susceptibility was evaluated after subtracting the core diamagnetic contribution estimated from Pascal's constants, $\chi_{core} = 2\chi_{core}(BETS) + \chi_{core}(I_2) + \chi_{core}(I^-)$. It is difficult to attribute the observed negative susceptibility solely to uncertainties in this correction. At present, the origin of the low-temperature susceptibility remains unclear, and further investigations will be required to establish its intrinsic behavior.}

\subsection{D. Site-selective spin susceptibilty}
The local spin susceptibility $\chi_\mu$ was derived through the form of $K_\mathrm{iso(aniso)}^\mu=A_\mathrm{iso(aniso)}\chi_\mu $using the hyperfine coupling constants determined above.
Figure~\ref{Fig7}(a) shows the temperature dependences of $\chi_\mathrm{A}$, $\chi_\mathrm{B}$, and $\chi_\mathrm{C}$.
The local susceptibilities exhibit a remarkable site dependence throughout the measured temperature range. The relation of $\chi_\mathrm{C} > \chi_\mathrm{A}>\chi_\mathrm{B}$ is maintained from ambient temperature down to 2~K. All three susceptibilities gradually decrease upon cooling.
The decrease becomes particularly pronounced below approximately 100~K and all of them approach zero at low temperatures. No discontinuity or anomaly indicative of magnetic order is observed. Figure~\ref{Fig7}(b) enlarges the low-temperature behavior of $\chi_\mathrm{B}$, which does not show the anomalous negative values as previously observed in $\alpha$-(ET)$_2$I$_3$. The different behaviors are discussed later.

\subsection{E. Nuclear spin-lattice relaxation rate}
Representative relaxation curves are shown in Figs.~\ref{Fig8}(a)--\ref{Fig8}(c).
Above 100~K, the relaxation curves are well fitted by the double-exponential function described in Sec.~II~C. The short relaxation rate is approximately one order of magnitude larger than the long one.
Upon cooling, the two relaxation rates gradually converge. Accordingly, the exponent $\beta$ of the stretched-exponential fit approaches unity down to 45~K (inset of Fig.~\ref{Fig8}(d)).
The effective $1/T_1$ determined from the stretched-exponential fitting is shown along with the $1/T_1$ of the two-component fitting in Fig.~\ref{Fig8}(d). It continuously decreases upon cooling and levels off at low temperatures below 20~K. Figure~\ref{Fig8}(e) presents an Arrhenius plot of $1/T_1$.
Fitting the high-temperature data to an activated form yields an activation energy of $\Delta/k_\mathrm{B} = 210 \pm 41$~K. No critical enhancement of the relaxation rate is observed around $T_\mathrm{MI}$. 
The saturation of $1/T_1$ at low temperatures is possibly due to impurity spins, consistent with the non-single exponential relaxation at low temperatures (Fig.~\ref{Fig8}(c)).

\section{IV. Discussion}
\subsection{A. Symmetry of the insulating phase}
The present NMR measurements provide direct microscopic information on the symmetry of the low-temperature insulating state of $\alpha$-(BETS)$_2$I$_3$.
As described in Sec.~II~B, the A and A$^\prime$ sites are related only by inversion symmetry and therefore produce equivalent NMR spectra in the absence of inversion-symmetry breaking.
Any symmetry-lowering transition would lift this equivalence and result in additional spectral splitting.
However, no such splitting is observed over the entire temperature range investigated. The spectra are consistently described by one quartet from the A (=A$^\prime$) site and two doublets from the B and C sites, demonstrating that the equivalence between the A and A$^\prime$ molecules is preserved across the metal–insulator transition.
This result is in sharp contrast to \alphaET at ambient pressure, where charge ordering breaks inversion symmetry and produces inequivalent molecular environments below the transition temperature.
Together with the previous X-ray diffraction measurements \cite{Kitou2021}, the polycrystalline $^{77}$Se NMR measurements \cite{Hiraki2011} and the infrared optical spectroscopy \cite{Savita2025} reporting no detectable symmetry lowering, the present NMR observations provide strong evidence against charge ordering as the origin of the insulating state of $\alpha$-(BETS)$_2$I$_3$.

The present data also place stringent constraints on the magnetic symmetry of the ground state.
The angular dependence of the spectra remains smooth and is accurately described by the single hyperfine-coupling tensor down to 2~K. No line splitting, anomalous broadening, or additional resonance features are observed.
Furthermore, $1/T_1$ exhibits no critical enhancement around the metal–insulator transition, which suggests no critical magnetic fluctuations even around TMI. Because NMR directly probes local magnetic fields at the nuclear sites, the absence of detectable internal fields indicates that the insulating phase remains essentially nonmagnetic down to at least 2~K.
We therefore conclude that the low-temperature insulating state of \alphaBETS preserves both inversion symmetry and time-reversal symmetry.

\subsection{B. Site-selective susceptibility and Dirac electronic states}
The site-selective spin susceptibilities provide insight into the underlying electronic structure. A characteristic feature of the present data is the relation, $\chi_\mathrm{C}>\chi_\mathrm{A}>\chi_\mathrm{B}$, which persists over the entire temperature range.
The same relation and similar temperature dependence of them was previously reported for \alphaET in its pressure-induced Dirac-fermion phase.
Theoretical studies have shown that this relation originates from the site-dependent contributions of molecular orbitals to different regions of a tilted Dirac cone in momentum space \cite{Hirata2016}.
The C-site molecular orbital predominantly contributes to the gentle-slope side of the cone, where the density of states is relatively large, whereas the B-site molecular orbital has large spectral weight on the steep-slope side characterized by a smaller density of states.
The A-site contribution is distributed more uniformly over momentum space and consequently exhibits an intermediate density of the states. The observation of the same relation in \alphaBETS therefore suggests that its low-energy electronic structure is likewise governed by tilted Dirac-cone bands.
This interpretation is consistent with recent first-principles calculations predicting Dirac-like electronic states near the Fermi level \cite{Tsumuraya2021}.
Although the relation resembles that observed in $\alpha$-(ET)$_2$I$_3$, quantitative differences are apparent. The overall magnitude of the local susceptibility is substantially reduced, indicating a lower density of states in $\alpha$-(BETS)$_2$I$_3$, which is consistent with increases in overlap integrals between molecular orbitals by the Se substitution.

\begin{figure*}
\centering
\includegraphics[width = 1.75\columnwidth]{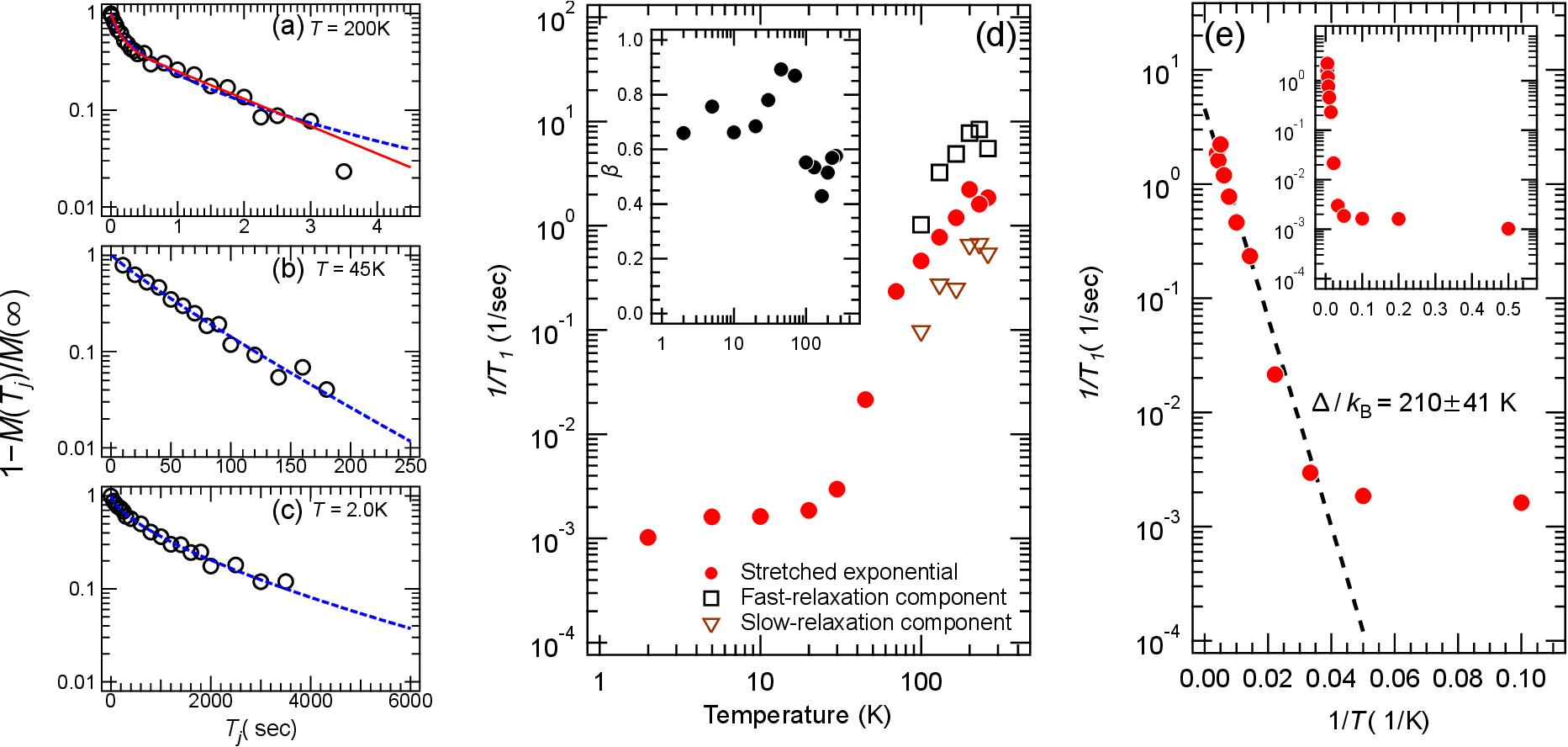}
\caption{\label{Fig8}(a)--(c) Relaxation curves under a field orientation of $ \theta= 31^\circ$ at 200~K, 45~K, and 2.0~K, respectively. Red curve shows the double-exponential fitting and blue dotted curves show the stretched-exponential fitting (see the text). (d) Temperature dependence of $1/T_1$. Square and triangle display $1/T_1$ estimated from short- and long-relaxation component from the double-exponential fittings. Red points display $1/T_1$ estimated from the stretched-exponential fittings. From those fittings, $\beta$ values are estimated as in inset. (e) Arrhenius plot of $1/T_1$ from the stretched exponential fittings, which is fitted by activation function with an activation energy of $\Delta/k_\mathrm{B} = 210\pm41$~K. Inset shows the plot in the low-temperature range, where $1/T_1$ level off.}
\end{figure*}

Particularly noteworthy is the low-temperature behavior of the B-site susceptibility. In $\alpha$-(ET)$_2$I$_3$, $\chi_\mathrm{B}$ changes sign below approximately 60~K as seen in Fig.~\ref{Fig8}(b). This behavior has been interpreted as a manifestation of ferrimagnetic-like spin excitations arising from inter-band Coulomb interactions around the Dirac point \cite{Hirata2016,Matsuno2017,Matsuno2018} In contrast, $\chi_\mathrm{B}$ in \alphaBETS shows no such behavior. The absence of a negative B-site susceptibility indicates that the inter-band Coulomb interactions may be weaker than $\alpha$-(ET)$_2$I$_3$. One possible interpretation is that \alphaBETS has a broader bandwidth, which is consistent with the relatively suppressed site-selective susceptibilities. In general, a larger bandwidth in a Dirac system leads to a smaller density of states around the Fermi energy. Another interpretation is that an energy gap opens at the Dirac points, as discussed in the next section.

\subsection{C. Indication of a spin--orbit-coupling-induced gap}
The site-selective susceptibility provides additional insight into the origin of the insulating state. Compared with $\alpha$-(ET)$_2$I$_3$, \alphaBETS exhibits a reduced spin susceptibility, which approaches zero below 40~K, and $1/T_1$ characterized by an activation energy of 210~K. Furthermore, ferrimagnetic-like spin excitations as observed in \alphaET is absent. All these observations indicate that the low-energy Dirac-fermion excitations are strongly depleted in the insulating phase. Such behavior is naturally expected when a gap opens around the Dirac point, suppressing the density of states available for low-energy excitations.
Among the mechanisms proposed thus far, the opening of a spin--orbit-coupling-induced gap provides the most natural explanation for the present observations. Because selenium atoms possess stronger spin--orbit coupling than sulfur atoms in $\alpha$-(ET)$_2$I$_3$, \alphaBETS has long been regarded as a promising candidate for a spin--orbit-coupling-driven insulator \cite{Ohki2020,Ohki2023}. A spin--orbit-coupling-induced gap naturally accounts for the suppression of the low-energy density of states while remaining fully compatible with the symmetry constraints established by the present NMR measurements. Although NMR alone cannot uniquely identify the microscopic origin of the gap, any viable theoretical description must simultaneously explain the preservation of inversion and time-reversal symmetries, the suppression of local spin susceptibility, and the absence of ferrimagnetic-like spin excitations.

\subsection{D. Constraits on correlated topological states}
Recent theoretical and experimental studies have proposed that \alphaBETS may host a correlated quantum spin Hall state stabilized by the interplay between spin--orbit coupling and electron correlations \cite{Ohki2022,Nomoto2023,Osada2023}.
The symmetry-preserving gapped Dirac state demonstrated by the present NMR measurements is compatible with this possibility. However, NMR primarily probes bulk electronic properties and is not directly sensitive to topological edge modes.
Therefore, the present data should not be regarded as direct evidence for a quantum spin Hall state. Rather, the present work offers the microscopic constraints that any candidate ground states must satisfy and suggests that a correlated quantum spin Hall phase is among possible ground states.
Its direct verification will require complementary probes sensitive to edge-state transport or spin-current phenomena.

\section{V. Conclusion}
We have performed site-selective $^{13}$C NMR measurements on single-crystalline \alphaBETS to investigate the nature of its low-temperature insulating state. The NMR spectra continuously vary through the metal--insulator transition and exhibit no additional splitting down to 2~K. The absence of spectral splitting demonstrates that inversion symmetry is preserved in the insulating phase. Furthermore, neither anomalous line broadening nor critical enhancement of $1/T_1$ is observed, indicating the absence of magnetic ordering with time-reversal-symmetry breaking.

By determining the hyperfine-coupling tensor, we estimated the site-specific spin susceptibilities. The susceptibilities have the relation of $\chi_\mathrm{C}>\chi_\mathrm{A}>\chi_\mathrm{B}$, which is a characteristic of tilted Dirac cones such as in \alphaET under pressure. In contrast, the B-site susceptibility that vanishes without its sign change indicates the absence of interaction-driven ferrimagnetic-like spin excitations as observed in $\alpha$-(ET)$_2$I$_3$.

These results establish that the insulating ground state of \alphaBETS preserves both inversion symmetry and time-reversal symmetry and is therefore distinct from conventional charge-ordered or magnetically ordered insulating states. Together with the observed suppression of low-energy spin excitations, the present findings are consistent with a symmetry-preserving gapped Dirac state due to the enhanced spin--orbit coupling and provide important microscopic constraints on theoretical descriptions of $\alpha$-(BETS)$_2$I$_3$, including proposed correlated topological phases. \alphaBETS should be a rare example of a molecular Dirac system whose insulating ground state preserves both inversion and time-reversal symmetries.

\section{ACKNOWLEDGMENTS}
This work was supported by the JSPS Grants-in-Aid for Scientific Research (Grant No. 17K05846, 18H05225, 19H01846, 20K20894, 20KK0060, 21K18144, 23K25815 and 24K21525), the Mitsubishi Foundation (Grant No. 202110014), Iketani Science and Technology Foundation (Grant No. 0381093-A)
and the Japan Science
and Technology Support for Pioneering Research Initiated
by the Next Generation (Grant No. JPMJSP2108).
K. K. acknowledges the support from the JSPS Core-to-Core Program (JPJSCCA20240001) and the Alexander von Humboldt Foundation. We also thank the Cryogenic Research Center at the University of Tokyo for supporting the low-temperature experiments.

\bibliography{BETS13Cnmr_ref.bib}
\end{document}